\documentclass[prl,twocolumn,showpacs,floatfix]{revtex4}

\usepackage{graphicx}
\usepackage{dcolumn}
\usepackage{bm}
\usepackage{amsmath,amsfonts,amssymb,amsthm}
\usepackage{braket}
\usepackage{slashed}
\usepackage{comment}
\usepackage{float}
\usepackage{tabularx}
\usepackage{times}
\usepackage{nicefrac}
\usepackage{epsf}
\usepackage{bbm}
\usepackage{color}
\usepackage{tikz}
\usepackage{amssymb}
\usepackage{graphicx}
\usepackage{verbatim}
\usepackage{txfonts}
\usepackage{dsfont}
\usepackage{enumitem}
\usepackage[utf8]{inputenc}
\usepackage{bbold}
\usepackage{mathtools}
\usepackage{lipsum, babel}
\usepackage[colorlinks=true,linkcolor=blue,citecolor=blue,urlcolor=blue]{hyperref}

\usepackage[normalem]{ulem}
\usepackage{siunitx}

\usepackage{verbatim}

\usepackage{pstricks,pst-node,pst-text,pst-3d}

\begin{document}
\thispagestyle{empty}
\title{Entanglement-induced enhancement in two-photon ionization
}
\title{Multiphoton ionization driven by quantum light
}

\title{ Entanglement enhanced multiphoton ionization }

\title{ Enhanced multiphoton ionization driven by quantum light }

%
%
\author{V. P. Kosheleva$^{1}$}
\email{valeriia.kosheleva@mpsd.mpg.de}
\author{S. Panahiyan$^{1,2}$}
\email{shahram.panahiyan@mpsd.mpg.de}
\author{A. Rubio$^{1,3}$}
\author{F. Schlawin$^{1,2,4}$}

\affiliation{$^1$  Max Planck Institute for the Structure and Dynamics of Matter, Center for Free Electron Laser Science, Luruper Chaussee 149, 22761 Hamburg, Germany}
\affiliation{$^2$ University of Hamburg, Luruper Chaussee 149, 22761 Hamburg, Germany}
\affiliation{$^3$ Center for Computational Quantum Physics (CCQ), The Flatiron Institute, 162 Fifth Avenue, New York, New York 10010, USA}
\affiliation{$^4$ The Hamburg Centre for Ultrafast Imaging, Luruper Chaussee 149, Hamburg D-22761, Germany}
%
\begin{abstract}

We present a framework for multiphoton ionization driven by arbitrary quantum states of light. Our simulations predict that cross sections can be enhanced by 
more than two orders of magnitude with momentum-entangled photons produced by modern nanoscale quantum light sources. 
The enhancement is tied to the broad angular spectrum of such sources, and 
is severely underestimated by conventional approaches using the paraxial approximation. 
Reasonable estimates of the resonant two-photon ionization cross section in sodium atoms indicate that these effects should be observable with current technology. 
\end{abstract}

    \maketitle
\begin{figure}
    \centering
    \includegraphics[width=0.8\linewidth]{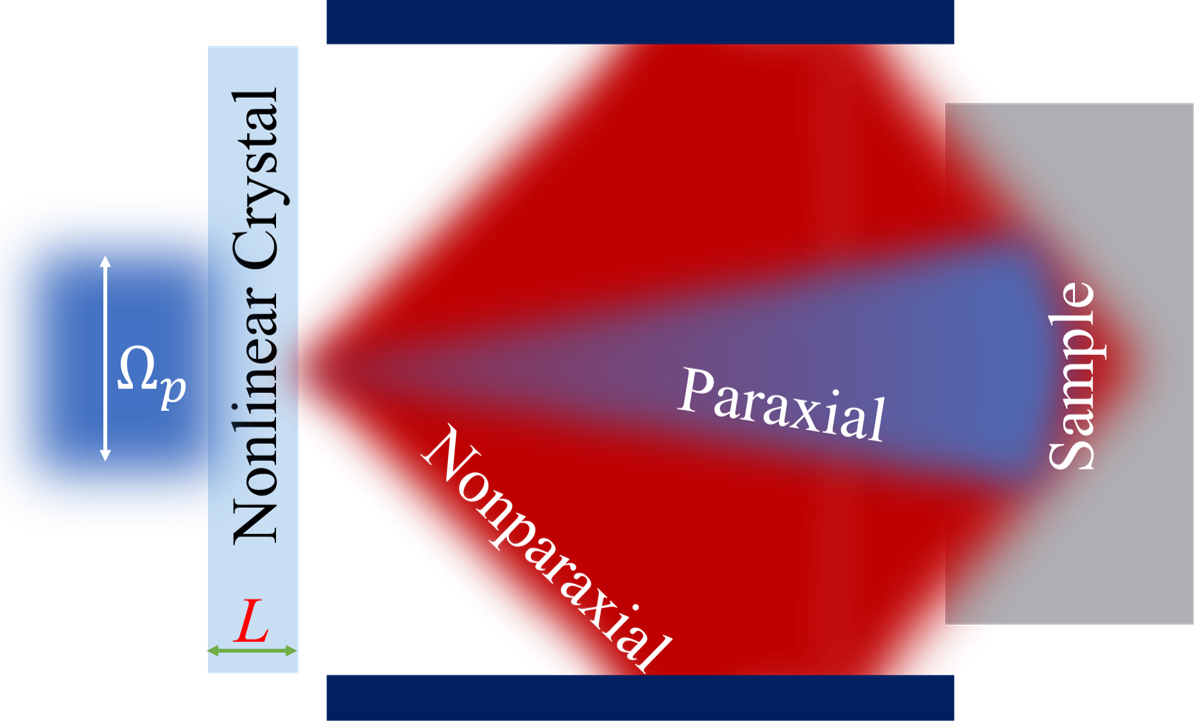}\llap{\parbox[b]{5.7in}{(a)\\\rule{0ex}{1.4in}}} \vspace{-1mm}   
    \includegraphics[width=0.99\linewidth]{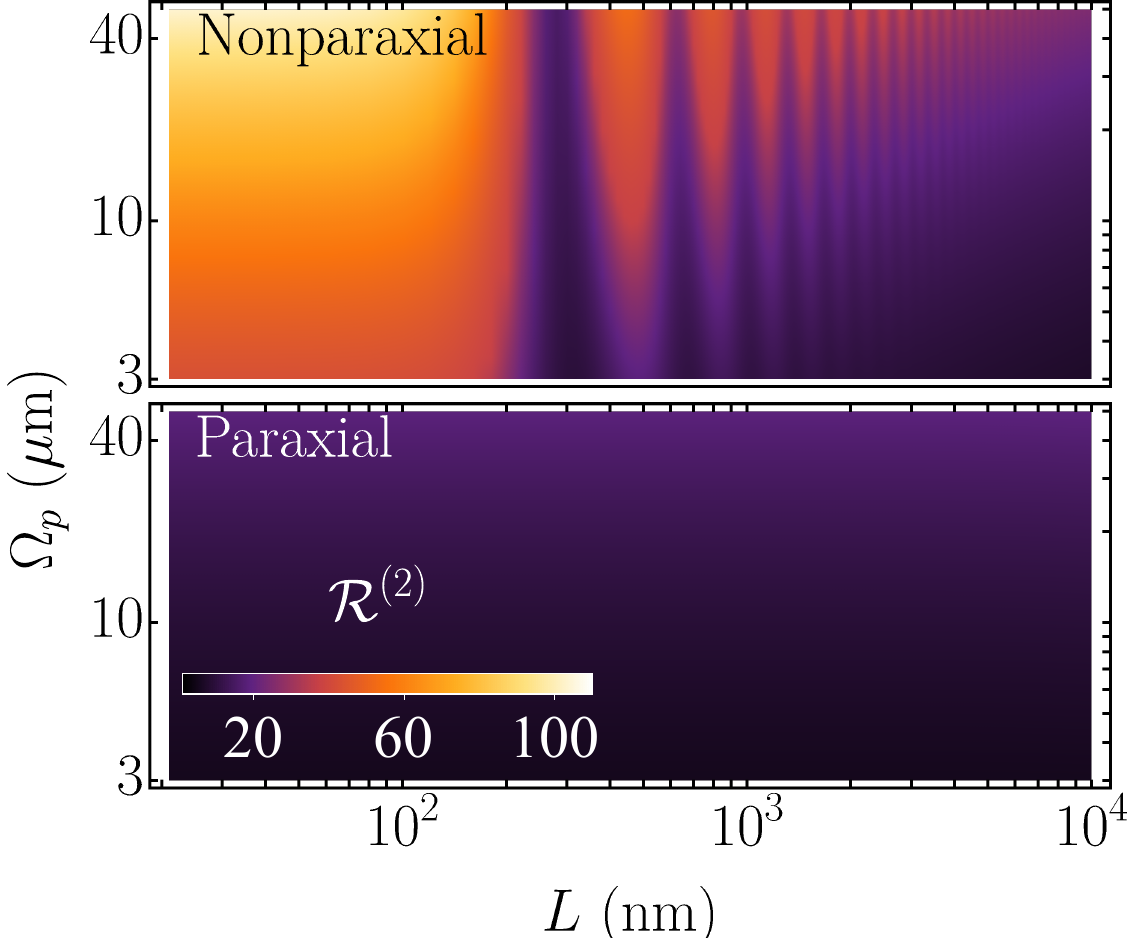} \llap{\parbox[b]{6.4in}{(b)\\\rule{0ex}{2.7in}}}
    \caption{(a) Schematic presentation of two-photon ionization of the sample with momentum entangled photons. A pump field 
    beam waist $\Omega_{p}$ interacts with a birefringent crystal with length $L$, where spontaneous parametric downconversion generates pairs of momentum-entangled photons. 
    These photon pairs 
    trigger resonant two-photon ionization in the sample. 
    The paraxial approximation accounts only for the narrow emission angles of entangled photons 
    and is valid for sufficiently large crystals, whereas short crystals allow much larger momentum mismatches and correspondingly broader angular spectra, where a careful treatment is necessary.
    (b) Enhancement ratio of the REMPI cross sections, \eqref{R_dip}, of entangled photons compared to two separable photons as a function of the pump beam waist $\Omega_p$ and length of the crystal $L$ for the ionization through dipole channel: (upper panel) nonparaxial approach and (lower panel) paraxial approximation. 
    We consider the central frequencies of signal and idler photons to match the energy of the ($3^2 S_{1 / 2}-4^2 P_{3 / 2}$) transition in a neutral sodium atom \cite{Sansonetti2008}. 
    The graphs are plotted on a logarithmic scale with base $10$.
    }
    \label{fig:schematic}
\end{figure}

%
%
%
%
%
%
%
%
\emph{Introduction.---} 
Nanoscale quantum light sources have emerged as novel, highly promising platforms for quantum information science and light-matter interactions. 
Recent experiments have demonstrated the generation of highly correlated photon pairs from single-layer crystals \cite{Santiago2022,Son2025,Fan2025}, 
subwavelength nonlinear films \cite{Santiago2021,SantosWeiss}, resonant metasurfaces \cite{Santiago2021b,Ding2023}, and other compact nonlinear structures \cite{Okoth2019}. These sources rely on parametric downconversion in a non-phase matched regime, where extremely broad bandwidth and momentum spreads can be generated (see Fig. \ref{fig:schematic}a).
These sources can be engineered to produce tunable polarization \cite{Sultanov22,Ma2023,Ma2025,Weissflog2024} or spatial \cite{Zhang2022} entanglement, support multiple entanglement links \cite{Santiago-Cruz2022}, and offer directional emission control \cite{son2023photon,weissflog2024directionally}. Such capabilities open new opportunities for probing matter with 
engineered spatio-temporal correlations. 
Research has only recently started to explore these opportunities for, e.g., linear quantum imaging~\cite{Vega2022}  but their wider ramifications for applications in sensing~\cite{Lawrie2019,degen2017quantum,pirandola2018advances}, imaging~\cite{Defienne2024,lemos2014quantum,lahiri2015theory,Vega2022,barreto2022quantum},  metrology \cite{Panahiyan2022,Panahiyan2023, pezze2018quantum,polino2020photonic,taylor2016quantum}, high-harmonic generation \cite{Rasputnyi2024, de-la-Pena2025}, and spectroscopy~\cite{Dorfman2016, Schlawin2018, Eshun2022} remain largely unexplored. 
\\
\indent
Investigating this potential requires careful consideration of the nonparaxial emission characteristics of nanoscale sources in these applications \cite{Okoth2019}. Multiphoton ionization, and its variant resonance-enhanced multiphoton ionization (REMPI), provides an experimentally accessible and highly sensitive platform for exploring this: its nonlinear nature is a direct probe of photon correlations, 
as recently demonstrated with bright squeezed light \cite{Heimerl2024}. 
Its strong sensitivity to intermediate-state resonances \cite{laforge2021resonance}, together with its intrinsic dependence on the spatial and correlation structure of the driving field, makes REMPI a clean and controllable setting to identify and quantify quantum advantages beyond the paraxial regime.
Beyond its role as a probe of nonclassical fields, REMPI underpins a broad range of applications, including plasma generation \cite{Sharma2018,Vagin2020}, chemical diagnostics \cite{RYSZKA2016,Sharma2020}, high-harmonic generation \cite{brown2012interference,ackermann2012resonantly,albar2025}, photoelectron spectroscopy \cite{kosheleva2020,Benda2021,laforge2021resonance}, ultracold atom–molecule dynamics and atom–atom collisions \cite{akerman2017trapping,Nichols2022,wolf2017state}, molecular chirality \cite{Bloch2021,vitanov2019highly,kastner2017intermediate,comby2018real}, catalytic surface chemistry \cite{shirhatti2018observation}, nuclear-spin conversion \cite{sugimoto2011electric}, and precision metrology \cite{Kohnke2024,germann2014observation,sinhal2020quantum}. 
%
\\
\indent
In this letter, we investigate REMPI driven by quantum states of light, taking the full spatial dependence of the photonic fields into account such that nonparaxial effects are captured properly. 
Our approach incorporates relativistic effects important for heavy atoms \cite{Kosheleva2022,Konecny2025,PhysRevA.103.042818,Glazov2019,PhysRevResearch.2.013364,Laskowski2010, Vidal2020, Kasper2020, Konecny2022,Hrobarik2011-NMR, Hrobarik2012, Vicha2016-2,Vicha2020,Martin2015,Bolvin2016,Misenkova2022,Komorovsky2023,Zinenko2023,Kosheleva202} and enables accurate predictions of angular distributions \cite{Sukhorukov_2012} and asymmetries \cite{Toffoli2002,Zapata2024} relevant to chiral discrimination \cite{PhysRevX.13.011044} and circular dichroism \cite{darquie2021valence}. Using an S-matrix formulation, we calculate perturbatively multiphoton ionization cross sections and, as a case study, predict that resonant two-photon ionization of the sodium atom by momentum-entangled photons may be enhanced by several orders of magnitude from momentum entanglement alone. 
This enhancement can be evaluated only by accounting for nonparaxial correlations, and almost vanishes in the paraxial limit. Moreover, our analysis shows that for dipole-mediated REMPI, the enhancement only depends on the photonic quantum state. This breaks down for higher multipole channels, pointing towards a new interesting light-matter interaction regime. 
%
%
%
 
%
\begin{figure*}
    \centering  
    \includegraphics[width=\linewidth]{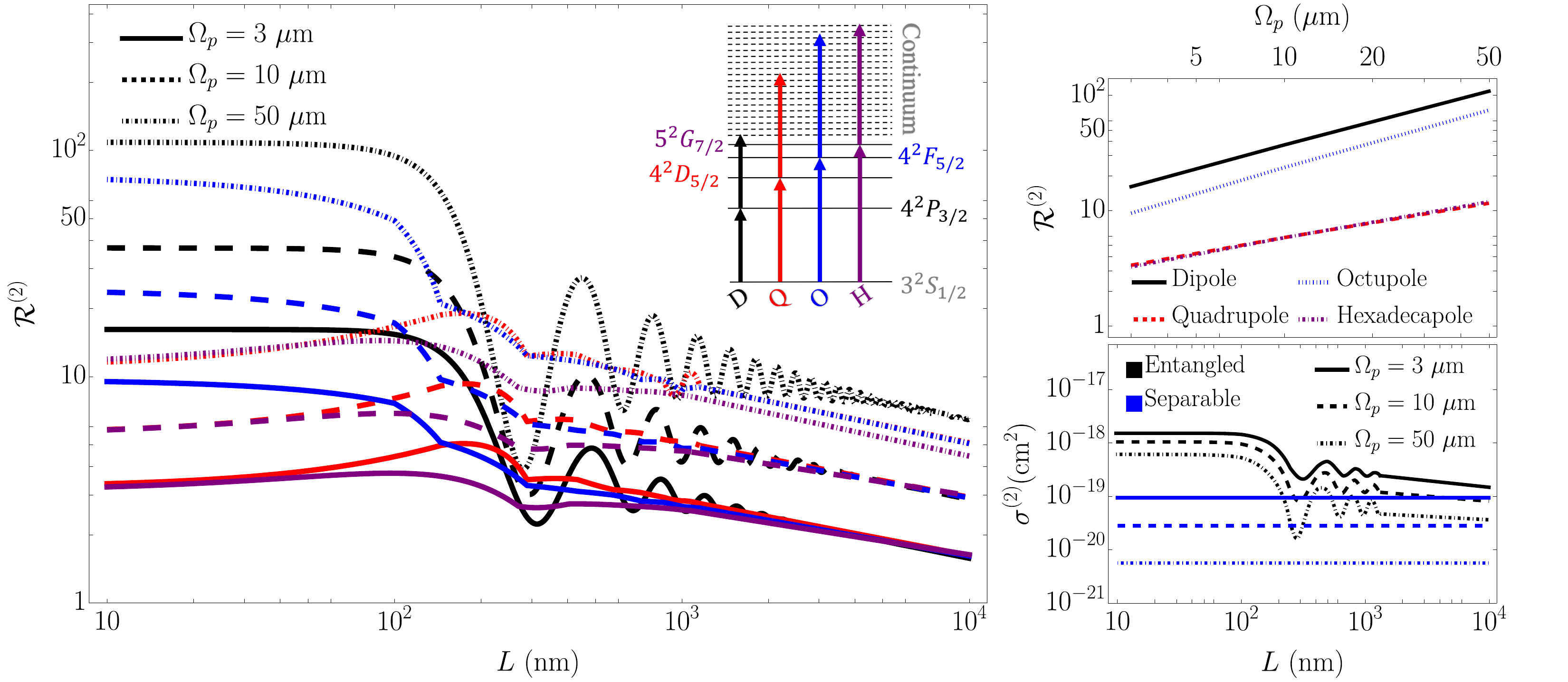}\llap{\parbox[b]{13.7in}{(a)\\\rule{0ex}{2.9in}}}\llap{\parbox[b]{4.9in}{(b)\\\rule{0ex}{2.6in}}}\llap{\parbox[b]{4.9in}{(c)\\\rule{0ex}{1.4in}}}
    \caption{    
    (a) The enhancement ratio of the REMPI cross sections ($\mathcal{R}^{(2)}$) \eqref{R} of a neutral sodium atom for dipole (black lines), quadrupole (red lines), octupole (blue lines), and hexadecapole (purple lines) as a function of the crystal length $L$ (which quantifies angular emissions of entangled photons). The considered beam waists of the pump field are $\Omega_{p}=3$ $\mu$m (solid lines), $\Omega_{p}=10$ $\mu$m (dashed lines), and $\Omega_{p}=50$ $\mu$m (dotted-dashed lines). (b) The enhancement ratio of the cross-section of neutral sodium atoms for dipole (solid black line), quadrupole (dashed red line), octupole (dotted-dashed blue line), and hexadecapole (dotted blue line) as a function of the 
     beam waists of the pump field ($\Omega_{p}$) for
     crystal length $L =10$ nm. (c) Total REMPI cross sections, $\sigma_{ent}^{(2)}$, with entangled (black lines) and separable (blue lines) photons via dipole channel as a function of crystal length $L$ for different beam waist. The transverse beam waist is \(\Omega_y = 50~\mu\mathrm{m}\), chosen large enough that the transverse momentum can be neglected. 
The pulse duration is \(T = 1.33~\mathrm{ps}\), corresponding to a photon bandwidth smaller than the \(4^2 P_{3/2}\!-\!4^2 P_{1/2}\) splitting, ensuring exclusive population of \(4^2 P_{3/2}\)
. All the graphs are plotted on a logarithmic scale with base 10.      
} 
    \label{fig:comparison}
\end{figure*} 
\emph{Total REMPI cross-section.---}
We model the process as the interaction between a quantum state of light \( |\psi_i\rangle \) and matter \( |\varphi_i\rangle \), initially prepared
in the state \( |\Psi_i\rangle = |\psi_i\rangle \otimes |\varphi_i\rangle \). After absorption of \( n \) photons, the system evolves into the total final state \( |\Psi_f\rangle = |\psi_f\rangle \otimes |\varphi_f\rangle \), where \( |\varphi_f\rangle \) describes the matter state comprising the photoelectron and the \emph{residual system} (e.g., an ion, molecular fragment, or ionized solid-state background), and \( |\psi_f\rangle \) is the final state of the light field.
\\
\indent
We define the total cross-section \( \sigma^{(n)}\) for \( n \)-photon ionization 
(see SM sec. A for details) 
\begin{align}
\sigma^{(n)} = 
\frac{\Gamma^{(n)}}{\phi^{\tilde{n}}},  
\label{total_cross_section}
\end{align}
where $\Gamma^{(n)}$ is the transition rate associated with the ionization process, and $\phi$ is a projection of the photon flux onto the axis perpendicular to the
plane of a target.
%
%
%
%
To account for photon bunching effects, the transition rate is divided by the $\tilde{n}$-th power of the photon flux, depending on the initial state of the field:
For example, for multiphoton absorption, entangled $n$-photon states generate a linear signal at low photon fluxes, such that $\tilde{n} = 1$~\cite{Perina1998,Fei1997}, whereas the signal of coherent pulses scales with the $n$-power of the photon flux, i.e. $\tilde{n} = n$. 

\emph{Photon flux.---} 
Without loss of generality, we assume the target is placed in the $xy$ plane; therefore, $\phi$ is the projection of the photon flux onto the $z$ axis. 
We define 
the photon flux of an 
arbitrary quantum state of light $\psi_{i}$ 
as
\begin{align}
\label{flux}
    \phi &= c\frac{\sum_{\lambda=1,2}\int_{-\infty}^{\infty} d{\bf k} \frac{k_z}{k} \langle  \psi_{i}| \hat{a}_{\lambda}^{\dagger}({\bf k})\hat{a}_{\lambda}({\bf k}) |\psi_{i}\rangle}{V_{\psi_{i}}}, 
\end{align}
where $c$ is the speed  of light in vacuum, $\hat{a}_{\lambda}({\bf k})$ is an annihilation operator of a photon with momentum ${\bf k}$ and polarization $\lambda$, $k_z$ is projection of ${\bf k}$ onto $z$ axis, $k = \left| {\bf k} \right|$, and $V_{\psi_{i}}$ is {\it effective volume} which quantifies the localization of the 
photonic state in space. 
This definition allows us to account for the strong wavepacket dispersion beyond the paraxial regime. 
It naturally incorporates the spatial and temporal confinement of realistic photon states, and 
Eq.~(\ref{total_cross_section}) reduces to the established semiclassical cross sections in the appropriate limits: 
Taking a plane wave as the initial state of light, the effective volume coincides with the full quantization volume since the state is completely delocalized \cite{Santra2009}. For a Gaussian pulse, the volume is determined by the transverse beam waist and longitudinal pulse length, and for entangled photons, it is further reduced due to correlations inherent in the quantum state, as we will discuss below.
\\
\indent
{ \emph{Enhancement ratio.---} To study the enhancement of the REMPI cross-section due to quantum correlations, we introduce the ratio between the REMPI cross sections induced by entangled light fields, $\sigma_{ent}$, and by suitable
separable states, $\tilde{\sigma}_{sep}$ (see discussion after Eq. \eqref{eq:twophotonent}), 
\begin{equation}
\mathcal{R}^{(n)} = \frac{\sigma_{ent}^{(n)}}{\tilde{\sigma}_{sep}^{(n)}}.
\label{R}
\end{equation}
Both cross sections are calculated from Eq.~\eqref{total_cross_section} with a linear photon flux dependence,  $\tilde{n}=1$, such that $\mathcal{R}^{(n)}$ is unitless. 
Crucially, if the ionization is 
resonant with intermediate dipole transitions and the field polarization is independent of the photons' propagation direction, $\mathcal{R}^{(n)}$ becomes independent of the material: For narrowband separable or entangled states with $n$-photon wavefunctions $F_{ent/sep}(\boldsymbol{k}_1,\cdots,\boldsymbol{k}_n)$, we find (for details, see SM sec. B\ref{sec:R_dipole})
\begin{align}
\mathcal{R}^{(n)} &=  \frac{V_{ent}}{V_{sep}}\frac{f_{ent}^{(n)}}{f_{sep}^{(n)}},
\label{R_dip}
\end{align}
where $V_{ent/sep}$ is the effective volume of the respective state, and $f_{ent/sep}^{(n)} = \frac{\left|\int_{-\infty}^{\infty} d \boldsymbol{k}_1 \cdots d \boldsymbol{k}_n F_{ent/sep}(\boldsymbol{k}_i,\cdots,\boldsymbol{k}_n)\right|^2}{ \int_{-\infty}^{\infty} d \boldsymbol{k}_1,\cdots, d \boldsymbol{k}_n 
|F_{ent/sep}(\boldsymbol{k}_1, \cdots,\boldsymbol{k}_n)|^2\left( \frac{k_{1,z}}{k_1}+\cdots+\frac{k_{n,z}}{k_n} \right)}$  quantifies the focusing of the $n$-photon wavefunction at the sample. 
Notably, the enhancement depends only on the light field, not on the sample. 
As we will demonstrate by simulations below, however, this result~\eqref{R_dip} breaks down when the quantum state of light is in resonance with a higher-order multipole transition. 
\\
\indent
\emph{Momentum-entangled photon pairs.---} 
We first illustrate this enhancement in the case of the cross section of two-photon ionization with momentum-entangled photons. 
%
More precisely, we consider a model of two narrowband entangled photons, both centered around frequency $\omega$, which are momentum-entangled in the $xz$ plane, i.e., the $y$ components of the momentum for both photons are very small, $k_{i/s,y}\approx0$, as investigated in~\cite{Vega2022}. 
This leads to $|\psi_{ent} \rangle = \int_{-\infty}^{\infty}\int_{-\infty}^{\infty} d \boldsymbol{k}_i d \boldsymbol{k}_s  F_{ent}\left(\boldsymbol{k}_i, \boldsymbol{k}_s\right) \hat{a}^{\dagger}_{\lambda_y}(\boldsymbol{k}_i)\hat{a}^{\dagger}_{\lambda_y}(\boldsymbol{k}_s)|0,0\rangle$ with  $\boldsymbol{k}_s$ and $\boldsymbol{k}_i$ being the momentum of signal and idler photons, respectively, and $\lambda_y$ being a linear polarization in the $y$ direction. 
The two-photon wavefunction is given by (for details, see SM sec. C\ref{sec:two_ph_ampl})
%
%
%
\begin{align}
F_{ent}\left(\boldsymbol{k}_i, \boldsymbol{k}_s\right) &=
\frac{C_{ent}}{\sqrt{2}} sinc\left(\frac{L\Delta k_z}{2}\right)E_p(\boldsymbol{k}_p, \Omega_p) \notag 
\\
&
\times
e^{-\frac{\Omega_y^2\left(k_{i,y}^2+k_{s,y}^2\right)-T^2\left(\left(\omega_i-\omega\right)^2+\left(\omega_s-\omega\right)^2\right)}{2}} \theta \left( k_{i,z}\right)\theta \left( k_{s,z}\right), \label{eq:twophotonent}
\end{align}
where $sinc$ is a sinc function, $L$ is the length of the crystal in $z$ direction, $\Delta k_z$ the momentum mismatch in $z$ direction, $\boldsymbol{k}_p$ and $\Omega_p$ are momentum and beam waist of the pump field $E_p$. The pulse duration is given by $T$, $\Omega_y$  characterizes the spread of the $k_{i/s,y}$ for the idler and signal photons, and $\theta \left( k_{i/s,z}\right)$ is Heaviside function, which enforces propagation in the positive $z$ direction. 
$C_{ent}$ is a normalization constant. 
The two-photon state is localized in the effective volume $ V_{ent} = C_{ent}$, and its degree of entanglement is determined by the length $L$ and the beam waist $\Omega_p$~\cite{WALBORN201087}. 
\\
\indent
We compare its REMPI cross section with a separable counterpart $F_{sep}$ which describes two narrowband Gaussian photons propagating along the $z$-axis, but features the same beam waist, pulse duration, central frequencies, and linear $y$-polarization as the signal and idler photons.
(see the supplementary material sec. C\ref{sec:two_ph_ampl} for explicit form of $F_{sep}$). With these two wave functions and Eq. \eqref{R_dip}, we can evaluate the enhancement without specifying the sample. The results are presented in Fig.~\ref{fig:schematic}(b).
The enhancement increases with the beam waist, and with decreasing crystal length but plateaus at very short lengths, relevant for nanoscale sources. 
In this regime, the down-converted signal and idler photons are emitted at large angles, resulting in substantial momentum uncertainty and stronger spatial correlations that drive the enhancement. 
At intermediate lengths, the enhancement oscillates with the crystal length, since the two-photon amplitude is proportional to a sinc function, as predicted, e.g., in entangled two-photon transparency~\cite{Fei1997}.
An enhancement of several orders of magnitude is shown to be possible from this analysis. This pronounced enhancement arises from 
the large longitudinal mismatch leading to a very broad angular spectrum of photons. 
%
%
%
\\
\indent
\emph{Paraxial vs nonparaxial treatment.---}
%
We repeat the same calculation using the paraxial approximation. The results for these calculations (paraxial and nonparaxial treatment) are plotted in Fig. \ref{fig:schematic}b. 
Strikingly, the paraxial approximation strongly underestimates the achievable enhancement, incorrectly suggesting that only an $\mathcal{O} (1)$ enhancement is possible in the depicted parameter range.
In addition, the paraxial approximation completely misses the oscillatory features in the intermediate-length crystal that we discussed above. 
Hence, it cannot describe the most interesting regime of quantum-enhanced REMPI. 

A distinct feature emerging only in this ultrathin-crystal regime is the formation of a plateau in  Fig. \ref{fig:comparison}a. 
This plateau arises because the sinc function governing longitudinal phase-matching in \eqref{eq:twophotonent} approaches unity, i.e. sinc$ (\Delta k L / 2) \simeq 1$. 
This is the non-phasematched regime which is characteristic of nanoscale entangled light sources with subwavelength crystal lengths~\cite{Okoth2019,Santiago2022,Son2025,Fan2025}. 
This saturation effect indicates an ultimate enhancement limit for momentum-entangled two-photon interactions, which has not been previously reported. 
It emerges in the nonlinear interactions considered here, and is absent, e.g., in linear quantum imaging~\cite{Vega2022}. 

\emph{REMPI beyond the dipole.}---
%
Given the results for the dipole channel, 
a natural question is whether higher-order multipole ionization channels can be enhanced in the same way. 
First, we note that Eq.~\eqref{R_dip} breaks down because the multipole moments matrix elements depend explicitly on the photonic wavevectors, such that photonic and sample correlation functions cannot be separated, and the potential enhancement now becomes explicitly sample-dependent. 
We thus consider a case study in which our sample is a neutral sodium atom and we model the process in the framework of a single electron approximation (see computational details \cite{salvat1995accurate, Kosheleva2018,Verolainen1982} in the supplementary material sec. A). 
We present the simulations of the enhancement ratios using Eq. \eqref{R}  for REMPI via dipole, quadrupole, octupole, and hexadecapole transitions in Fig. \ref{fig:comparison}. For ionization via the dipole channel, the results obtained from Eqs. \eqref{R} and \eqref{R_dip} coincide with very high accuracy (See Fig. A1 in SM).

In the dipole channel, the most significant enhancement in REMPI occurs for ultrathin crystals ($L<300$ nm),
where non-paraxial effects are the most pronounced (see Fig. \ref{fig:comparison}a). 
This remains true in the octupole channel, but surprisingly the enhancement in both the quadrupole and hexadecapole channels peaks 
at a finite crystal length, 
approximately given by the
wavelength of the quadrupole (hexadecapole) transition, respectively. 
In any channel, as the crystal length increases, moving toward the paraxial regime, the enhancement of the cross section decreases and vanishes for sufficiently large crystal lengths. 
The dipole transition exhibits the most pronounced enhancement in the cross-section, followed by the octupole (see Fig. \ref{fig:comparison}a and \ref{fig:comparison}b). 
As we already discussed for the ionization via the dipole channel, the former enhancement ratio is determined solely by the light properties. 
In contrast, the quadrupole and hexadecapole transitions display significantly weaker enhancement across the depicted parameter ranges. 

In all channels, the enhancement increases monotonically with beam waist, as the excitation probability of separable states naturally decreases with beam waist (i.e. with focusing). 
However, the rate of increase is channel-dependent (Fig. \ref{fig:comparison}b): the octupole and hexadecapole transitions show steeper growth compared to the dipole and quadrupole cases. 
Consequently, there is a crossover for very large beam waists, where the enhancement of 
higher-order multipole channels (octupole and hexadecapole) can surpass their lower-order counterparts (dipole and quadrupole). 
This can be observed in Fig. \ref{fig:comparison}b, where the hexadecapole transition eventually exceeds the quadrupole in enhancement for sufficiently large beam waists. Overall, the results presented in Fig. \ref{fig:comparison}a and b clearly confirm that Eq.~\eqref{R_dip} breaks down when other ionization channels are considered, and the enhancement depends on the material explicitly. 
\\
\indent
\emph{Symmetry of the ionization channels.---}
The observed variation in enhancement originates from the interplay between the photonic wavefunction and the material response. For the quadrupole and octupole ionization channels, the photon fluxes are nearly identical, since the corresponding frequencies $\omega$ are almost the same in both cases. The enhancement therefore cannot be attributed to flux differences alone. Instead, it arises from the overlap between the photonic wavefunction $F_{ent}$ and the material response in $\mathbf{k}$-space, which is larger for the octupole (even-parity) channel than for the quadrupole (odd-parity) one. A similar trend holds for other channels: the enhancement ratio $\mathcal{R}^{(2)}$ is maximal for the dipole (odd-parity) channel, reflecting the largest $\mathbf{k}$-space overlap, as the material response effectively decouples from $F_{ent}$.

Consequently, at short crystal lengths, where the angular range for signal and idler photons is broad, odd multipole transitions
with parity-favored pathways (dipole and octupole) exhibit
significantly enhanced coupling, resulting in the highest enhancement.
As the crystal length increases, the angular range for signal and idler photons narrows, reducing
the distinctiveness between channels and diminishing the
enhancement differences in any excitation channel. 

\emph{Quantum-enhanced REMPI cross section.---}
We can express the entangled two-photon ionization cross section $\sigma_{ent}^{(2)}$ in terms of the classical cross section, using $\sigma_{cl}^{(2)} = \frac{\tilde{\sigma}_{sep}^{(2)}}{\phi_{sep}}$. This yields
\begin{align}
 \sigma_{ent}^{(2)}  = \mathcal{R}^{(2)} \phi_{sep}\sigma_{cl}^{(2)}, \label{eq.relation}
    \end{align}
where the material properties are encoded in the classical cross section $\sigma_{cl}^{(2)}$ \cite{shabaev:2002:119,Kosheleva2022}. The classical cross section is weighted by the ratio $\mathcal{R}^{(2)}$ which quantifies the momentum 
correlations of the entangled photon pair. 
We note an important distinction between Eq.~(\ref{eq.relation}) and similar expressions in the literature: 
Earlier expressions for entangled cross sections accounted for quantum correlations in momentum and energy separately 
~\cite{Fei1997}, such that the factor $ \mathcal{R}^{(2)} \phi_{sep}$ can be replaced by the inverse of the so-called entanglement area and time. 
Eq.~\eqref{eq.relation} instead allows for arbitrary spatial profiles and propagation directions, and can account, e.g, for spatio-temporal entanglement~\cite{PhysRevLett.97.243903,Dickinson2025}. 

In Fig.~\ref{fig:comparison}c, we show estimates of the total REMPI cross section for entangled and separable photons via the dipole channel as a function of the crystal length and different beam waists of a pump beam $\Omega_p$. To this end, we have to choose realistic values for the pulse duration and the transverse beam waist (see Fig.~\ref{fig:comparison}c). 
%
}
The cross sections up to $\lesssim 10^{-18}$~cm$^2$ are on par with the largest reported entangled two-photon absorption cross sections in molecules~\cite{Villabona-Monsalve2017,Eshun2022} or with phasematched sum frequency generation~\cite{Dickinson2025, Landes2024}, and orders of magnitude larger than those measured in competing experiments~\cite{ Landes2024, Tabakaev2022, Parzuchowski2021, He2024, Corona-Aquino2022, pandya2024robustdetectionentangledtwophoton}. 
The reason lies, of course, in the resonant transitions considered here which increase the cross sections dramatically. This is enabled by the detection of photoelectrons, where the linear signal cannot be spoiled by other processes as is the case in a purely optical setting~\cite{Mikhaylov2022}. Hence, our simulations establish REMPI as an attractive platform to explore entangled two-photon physics and for entanglement-enhanced sensing, spectroscopy and imaging. 
\\
\indent
\emph{Conclusion.---}  
We have investigated resonant multiphoton ionization driven by entangled light produced by nanoscale sources. 
By describing the ionization processes beyond the paraxial approximation, we uncovered a large enhancement of the ionization cross section. We further found channel-dependent enhancement that are entirely missed in paraxial models. These results demonstrate that the metrological advantages of entangled light emerge most strongly in the beyond-paraxial regime, where large momentum  correlations determine the ionization efficiency. This insight highlights the relevance of emerging nonclassical light sources based on ultrathin crystals, metasurfaces, and subwavelength nonlinear films, which naturally operate in this regime. Beyond revealing this effect, our framework provides a general foundation for exploring quantum-light-driven ionization, chiral and symmetry-sensitive photoresponses, and precision spectroscopy with modern nanoscale light sources.

We close by mentioning that we only analyzed the impact of momentum correlations in this paper. Extending the analysis to higher-dimensional spatio-temporal entanglement~\cite{Gatti2009} will likely enhance the total cross section even further. 
Although our case study focuses on neutral sodium atoms, the perturbative S matrix formalism is fully general and applies to both finite systems, such as molecules or other atomic species, and extended periodic systems, including solids.

\begin{acknowledgments} 
S. P. acknowledges support from the Hamburg Quantum Computing Initiative (HQIC) project EFRE. The project is co-financed by ERDF of the European Union and by "Fonds of the Hamburg Ministry of Science, Research, Equalities and Districts (BWFGB)". F. S. acknowledges support from 
the research unit "FOR5750: OPTIMAL" - project ID 531215165. This work was also supported by the European Research Council (ERC-2024-SyG- 101167294 ; UnMySt) and the Cluster of Excellence "Advanced Imaging of Matter" of the Deutsche Forschungsgemeinschaft (DFG) - EXC 2056 - project ID 390715994.
A. R. acknowledges support from the Max Planck-New York City Center for
Non-Equilibrium Quantum Phenomena. The Flatiron Institute is a division of the Simons Foundation.
\end{acknowledgments}
%
%
%
%
%
%
%
\bibliography{references}
%
%

\newpage

\begin{widetext}
\renewcommand{\theequation}{A.\arabic{equation}}
\setcounter{equation}{0}

\section{Supplementary material}
SI units are used throughout the paper, unless stated otherwise.

\renewcommand{\thefigure}{A\arabic{figure}}
\setcounter{figure}{0}

\section{A. Definition of the cross section for two-photon ionization with quantum states of light
}
\label{sec:total_cross_section}
The total cross-section $\sigma^{(n)}$ is given by
\begin{align}
\sigma^{(n)} = \frac{\Gamma^{(n)}}{\phi^{\tilde{n}}} =\frac{1}{g_i} \sum_{i}\sum_f
\frac{|S^{(n)}_{if}|^2}{T\phi^{\tilde{n}}},  
\label{total_cross_section-G}
\end{align}
where $T$ is the interaction time (pulse duration), $S^{(n)}_{if}$ is the transition matrix element from the total initial state \( |\Psi_i\rangle \) to the total final state  \( |\Psi_f\rangle \).
We sum over the final  states ($\sum_{f}$), and average over the initial degenerate states ( $\frac{1}{g_i}\sum_{i}$) with $g_i$ being a degeneracy of the initial state.
All quantum numbers required to characterize
the initial state of the system, $|\Psi_{i}\rangle$, (and also final state $|\Psi_{f}\rangle$) are specified by the multi-index $i$ ($f$). We note that in the case when these quantum numbers are continuous, the sums in Eq. \eqref{total_cross_section-G} must be interpreted as integrals with proper normalizations.
\\
\indent
For the case of the ionization by two-entangled photons, the Eq. \eqref{total_cross_section-G} reduces to:
\begin{align}
\sigma^{(2)} =\frac{\Gamma^{(2)}}{\phi^{\tilde{n}}}  =  \frac{\frac{1}{g_i}\sum_i\sum_{\mu_f}\int_{-\infty}^{\infty} d\boldsymbol{p}_f\left|S^{(2)}_{i\mu_f}\left(\boldsymbol{p}_f\right)\right|^2}{T\phi^{\tilde{n}}},  \label{eq.Sigma}
\end{align}
where $S^{(2)}_{i\mu_f}\left(\boldsymbol{p}_f\right)$ is a transition matrix element from the total initial state given by \( |\Psi_i\rangle = |\psi_i\rangle \otimes |\varphi_i\rangle \) with $|\varphi_i\rangle $ being an initial electronic ground state to the total final state \( |\Psi_f\rangle = |0,0\rangle \otimes |\varphi_f\rangle \) with $|\varphi_f \rangle = |\varphi_{\rm{res}}, \boldsymbol{p}_f \mu_f\rangle $ being a wave function of the final state of a system comprising the residual system state $|\varphi_{\rm{res}}\rangle $ and photoelectron wave function $|\boldsymbol{p}_f \mu_f\rangle$. Here, the photoelectron is characterized by helicity $\mu_f$ and definite asymptotic momentum $\boldsymbol{p}_f = (p_f,\theta_f,\phi_f)$ with $p_f = |\boldsymbol{p}_f|$ and $\theta_f$ and $\phi_f$ being polar and azimuthal angles, respectively. 
The initial photonic state $|\psi_i\rangle$ has general form,
\begin{align}
|\psi_i\rangle  =  \int_{-\infty}^{\infty}\int_{-\infty}^{\infty} d \boldsymbol{k}_i d \boldsymbol{k}_s F\left(\boldsymbol{k}_i, \boldsymbol{k}_s\right) \hat{a}^{\dagger}_{\lambda_y}(\boldsymbol{k}_i)\hat{a}^{\dagger}_{\lambda_y}(\boldsymbol{k}_s)|0,0\rangle, 
\end{align}
where $F\left(\boldsymbol{k}_i, \boldsymbol{k}_s\right) = F\left(\boldsymbol{k}_s, \boldsymbol{k}_i\right) = \frac{C_0}{\sqrt{2}} \tilde{F}\left(\boldsymbol{k}_i, \boldsymbol{k}_s\right)$ with $C_0 = \sqrt{\frac{1}{\int d\boldsymbol{k}_i \int d\boldsymbol{k}_s\left|\tilde{F}(\boldsymbol{k}_i,\boldsymbol{k}_s)\right|^2}}$ being a normalization factor.
The transition matrix element is given by 
\begin{align}
S^{(2)}_{i\mu_f}\left(\boldsymbol{p}_f\right) &= -\frac{  2\pi i e^2 c^2}{\hbar^2}  \int_{-\infty}^{\infty}\int_{-\infty}^{\infty} d \boldsymbol{k}_i d \boldsymbol{k}_s F\left(\boldsymbol{k}_i,\boldsymbol{k}_s\right)\bigg[  \sum_{n}\frac{\left\langle\varphi_{\rm{res}}, \boldsymbol{p}_f \mu_f\left|\hat{R}_{\boldsymbol{k}_s,\lambda_y}\right| \varphi_n\right\rangle\left\langle\varphi_n\left|\hat{R}_{ \boldsymbol{k}_i,\lambda_y}\right| \varphi_{i}\right\rangle}{E_f/\hbar-\omega_{s}-E_n/\hbar\left(1- \frac{i\Gamma_n}{2}\right)} 
+   \notag
\\
& +\sum_{n}\frac{\left\langle\varphi_{\rm{res}}, \boldsymbol{p}_f \mu_f\left|\hat{R}_{\boldsymbol{k}_i,\lambda_y}\right| \varphi_n\right\rangle\left\langle\varphi_n\left|\hat{R}_{\boldsymbol{k}_s,\lambda_y}\right| \varphi_{i}\right\rangle}{E_f/\hbar-\omega_{i}-E_n/\hbar\left(1- \frac{i\Gamma_n}{2}\right)}\bigg]
\delta(E_f/\hbar-\omega_{i}-E_i/\hbar-\omega_{s}), \label{eq.S2}
\end{align}
in which $\Gamma_n$ and $E_n$ are a lifetime and energy of the intermediate state $\varphi_n$, respectively, $E_i$ is the energy of the initial state $\varphi_i$ and $E_f = \varepsilon_f+E_{{\rm res}}$ with $\varepsilon_f$ and $E_{{\rm res}}$ being an energy of photoelectron and residual system, respectivly. Here, the interaction operator $\hat{R}_{\boldsymbol{k},\lambda} = \sum_{i}^{N_e} \boldsymbol{\alpha}_i \cdot {\bf A}_{\boldsymbol{k}, \lambda}({\bf r}_i)$ where $N_e$ is a number of the electrons in a system and the amplitudes ${\bf A}_{\boldsymbol{k}, \lambda}({\bf r}_i)$ are given by ${\bf A}_{\boldsymbol{k}, \lambda}({\bf r}_i)=
\sqrt{\frac{ \hbar }{2 \omega_{\boldsymbol{k}}\epsilon_0(2 \pi)^3}}
e^{\mathrm{i} \boldsymbol{k}\cdot \boldsymbol{r}_i} \boldsymbol{e}_{\lambda_y}$.
In the above formula, we will replace $F\left(\boldsymbol{k}_s,\boldsymbol{k}_i\right)$ with Eqs. \eqref{eq.FENT} and \eqref{eq.FSEP} in the following, depending on whether ionization takes place with entangled or separable states, respectively. Since we consider the case of the narrowband photons, we substitute $\omega_i=\omega_s=\omega$.
We can remove the dependency on interaction time $T$ for sufficiently long pulses in Eq. \eqref{eq.Sigma}, via
using the following relation for the  delta function 
\begin{equation}
\delta^2(E_f/\hbar-2\omega-E_i/\hbar)=\lim _{T \rightarrow \infty} \delta(E_f/\hbar-2\omega-E_i/\hbar) \int_{-T / 2}^{T / 2} \frac{d t}{2 \pi} \mathrm{e}^{\mathrm{i}(E_f/\hbar-2\omega-E_i/\hbar) t}=\lim _{T \rightarrow \infty} \delta\left(\left(\varepsilon_f+E_{{\rm res}}\right)/\hbar-2\omega-E_i/\hbar\right) \frac{T}{2 \pi}. \label{eq.delta}
\end{equation}
By inserting it in Eq. \eqref{eq.S2}, we can cancel out $T$-dependence in the total cross section. Using the identity $d \boldsymbol{p}_f = \frac{1}{c^2} p_f \varepsilon_f d \varepsilon_f d \Omega_f$ (in which $d\Omega_f = sin\theta_f d\theta_f d\phi_f$ denotes the solid angle) and integrating over $d\varepsilon_f$, we get the expression for total cross section as
\begin{align}
\sigma^{(2)} =  
 \frac{\hbar p_f \varepsilon_f}{2\pi c^2}\frac{1}{g_i}\sum_i\sum_{\mu_f}\int d\Omega_f \frac{\left| \tau_{i\mu_f}^{(2)}\left( \boldsymbol{p}_f\right) \right|^2}{\phi^{\tilde{n}}},
 \label{sigma_2}
\end{align}
where the transition amplitude $\tau_{i\mu_f}^{(2)}\left( \boldsymbol{p}_f\right)$ is given by
\begin{align}
\tau_{i\mu_f}^{(2)}\left( \boldsymbol{p}_f\right) &= -\frac{  8\pi e^2 c^2}{\hbar^2}   \frac{1}{i\Gamma_n}\sum_n\int_{-\infty}^{\infty}\int_{-\infty}^{\infty} d \boldsymbol{k}_i d \boldsymbol{k}_s F\left(\boldsymbol{k}_i,\boldsymbol{k}_s\right)  \left\langle\varphi_{\rm{res}}, \boldsymbol{p}_f \mu_f\left|\hat{R}_{\boldsymbol{k}_s,\lambda_y}\right| \varphi_n\right\rangle\left\langle\varphi_n\left|\hat{R}_{ \boldsymbol{k}_i,\lambda_y}\right| \varphi_{i}\right\rangle. \label{eq.TauF}
\end{align}
Here $\varepsilon_f = E_i + 2\omega - E_{{\rm res}}$ and we consider the case of the resonant ionization, when the intermediate state has the energy $E_n = E_i+\omega$. 
Finally, using Eq. \eqref{flux}, the photon flux will be
\begin{align}
    \phi &= c\frac{\sum_{\lambda=1,2}\int_{-\infty}^{\infty} d{\bf k} \frac{{\bf k}\cdot {\bf e}_z}{k} \langle  \psi_{i}| \hat{a}_{\lambda}^{\dagger}({\bf k})\hat{a}_{\lambda}({\bf k}) |\psi_{i}\rangle}{V_{\psi_{i}}} = \frac{2c}{V_{\psi_i}}\int_{-\infty}^{\infty}\int_{-\infty}^{\infty} d \boldsymbol{k}_i d \boldsymbol{k}_s \left| F\left(\boldsymbol{k}_s,\boldsymbol{k}_i\right) \right|^2 \left( \frac{k_{i,z}}{k_i}+\frac{k_{s,z}}{k_s} \right), 
    \label{flux_sm}
\end{align}
where we used the fact that $V_{\psi_i} = C_0$ from unit analysis.
Now, to obtain $\sigma^{(2)}_{ent}$, one should use Eqs. \eqref{eq.delta}-\eqref{eq.TauF} with $F\left(\boldsymbol{k}_s,\boldsymbol{k}_i\right)$ given by Eq. \eqref{eq.FENT}, $C_0 = C_{ent}$, and $\tilde{n} = 1$. To obtain $\tilde{\sigma}^{(2)}_{sep}$, we use Eqs. \eqref{eq.delta}-\eqref{eq.TauF} with $F\left(\boldsymbol{k}_s,\boldsymbol{k}_i\right)$ given by Eq. \eqref{eq.FSEP}, $C_0 = C_{sep}$, and $\tilde{n} = 1$.
\\
\indent
\emph{Single active electron approximation in sodium atom.---}
In the present work, we consider two-photon ionization of a valence electron of a neutral sodium $(Z=11)$ atom.
We describe this process within the framework of the single-active-electron approximation (SAE). 
The active electron in the ground $3 s$ ($|\varphi_i \rangle$), excited,($|\varphi_n \rangle$), and continuum $|\boldsymbol{p}_f \mu_f \rangle$ states are described by the wave functions being the solutions of the Dirac equation with the effective potential describing the electric field of the nucleus and the spectator electrons. We chose the $z$ axis to be the quantization axis.
Then the amplitude reduces to
\begin{align}
\tau_{m_i\mu_f}^{(2)}\left( \boldsymbol{p}_f\right) &= -\frac{  8\pi e^2 c^2}{\hbar^2}   \frac{1}{i\Gamma_n}\sum_{m_n}\int_{-\infty}^{\infty}\int_{-\infty}^{\infty} d \boldsymbol{k}_i d \boldsymbol{k}_s F\left(\boldsymbol{k}_i,\boldsymbol{k}_s\right)  \left\langle\varphi_{f}\left|\boldsymbol{\alpha} \cdot \boldsymbol{A}_{\boldsymbol{k}_s,\lambda_y}\right| \varphi_n\right\rangle\left\langle\varphi_n\left|\boldsymbol{\alpha} \cdot \boldsymbol{A}_{ \boldsymbol{k}_i,\lambda_y}\right| \varphi_{i}\right\rangle, \label{eq.TauFNa}
\end{align}
where we sum over the projections of total angular momentum onto the quantization axis, $m_n$.
The wave function of the photoelectron $| \boldsymbol{p}_f \mu_f \rangle$ has the following form \cite{kosheleva2020},
\begin{equation}
| \boldsymbol{p}_f \mu_f \rangle = \sqrt{\frac{c^2}{4 \pi  \varepsilon_f p_f} }\sum_{\kappa m_j} C_{l 0}^{j \mu_f} 1_{1 / 2 \mu_f} i^l \sqrt{2 l+1} e^{-i \delta_k} D_{m_j \mu_f}^j\left(\phi_f, \theta_f, 0\right) \Phi_{\varepsilon \kappa m_j}(\boldsymbol{r}).
\label{eq:free_wf}
\end{equation}
Here $\kappa=(-1)^{l+j+1 / 2}(j+1 / 2)$ is the Dirac quantum number determined by the total angular momentum $j$ and the parity $l$, $\delta_\kappa$ is the phase shift induced by the scattering potential, and $\Phi_{\varepsilon \kappa m_j}(\boldsymbol{r})$ is the partial-wave solution of the Dirac equation in the scattering field \cite{salvat1995accurate, Kosheleva2018}.
And the total cross section is given by
\begin{align}
\sigma^{(2)} =  
 \frac{\hbar p_f \varepsilon_f}{2\pi c^2}\frac{1}{2j_i+1}\sum_i\sum_{\mu_f}\int d\Omega_f \frac{\left| \tau_{m_i\mu_f}^{(2)}\left( \boldsymbol{p}_f\right) \right|^2}{\phi^{\tilde{n}}},
 \label{sigma_2Na}
\end{align}
where $\varepsilon_f = E_i + 2\omega$ and we consider the case of the resonant ionization, when the intermediate state has the energy $E_n = E_i+\omega$. 
Here, the averaging over the initial degenerate states is expressed explicitly as $\frac{1}{g_i}\sum_i = \frac{1}{2j_i+1}\sum_{m_i}$, where $j_i$ is a total angular momentum of the initial electronic state $\varphi_i$ and $m_i$ its projection onto quantization ($z$ axis).
\\
\indent
We utilize the so-called $X \alpha$ central potential whose parameters are adjusted in such a way as to reproduce the energy of dipole ($3^2 S_{1 / 2}-4^2 P_{3 / 2}$), quadrupole ($3^2 S_{1 / 2}-4^2 D_{5 / 2}$), octupole ($3^2 S_{1 / 2}-4^2 F_{5 / 2}$), and hexadecapole  ($3^2 S_{1 / 2}-5^2 G_{7 / 2}$) transitions, respectively, namely $3.753 \, 293$ eV, $4.283\,461$ eV, $4.288\,194$ eV, and $4.594\,759$ eV~\cite{Sansonetti2008}. 
The corresponding lifetimes $\Gamma_n$ of the intermediate $4^2 P_{3 / 2}$, $4^2 D_{5 / 2}$, $4^2 F_{5 / 2}$, and $5^2 G_{7 / 2}$ are 90 ns, 55 ns, 72 ns, and 235 ns, respectively \cite{Verolainen1982}. The radial Dirac equation with the effective potential is solved by the modified radial package \cite{salvat1995accurate}.
\begin{figure*}
    \centering  
    \includegraphics[width=0.4\textwidth]{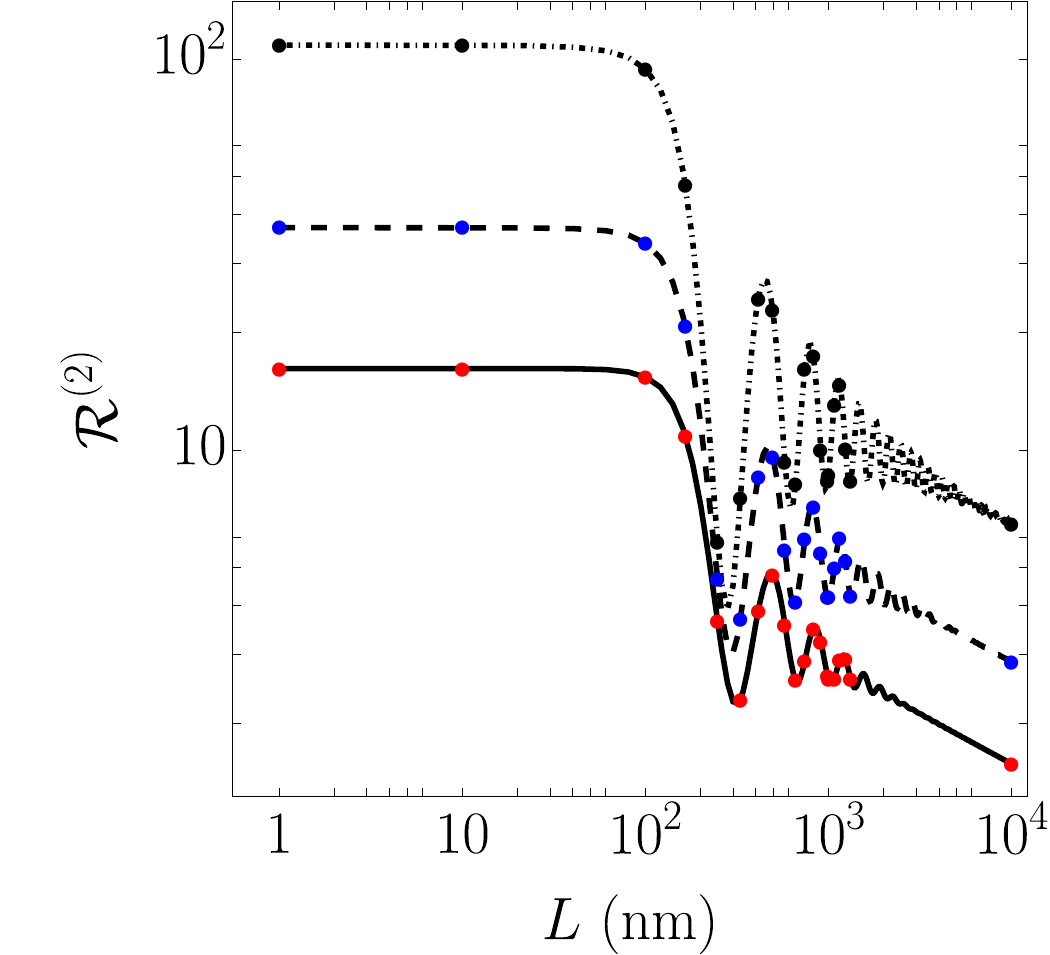}  
    \caption{    
    Enhancement ratio of the dipole channel as a function of the crystal length for three different beam waists: $\Omega_p = 3~\mu\text{m}$ (solid line), $\Omega_p = 10~\mu\text{m}$ (dashed line), and $\Omega_p = 50~\mu\text{m}$ (dash-dotted line). The curves are obtained from Eq.~\eqref{R_dip}, while the discrete markers represent the numerical evaluation of Eq.~\eqref{R}, showing excellent agreement between analytical and numerical results.} 
    \label{Fig_SM}
\end{figure*}
\section{B. Calculation of enhancement ratio: the dipole channel}
\label{sec:R_dipole}
When considering the transition amplitude $\tau_{i\mu_f}^{(2)}\left( \boldsymbol{p}_f\right)$ through the dipole channel in a dipole approximation, we can approximate the amplitude  by ${\bf A}_{\boldsymbol{k}, \lambda}({\bf r}_i) \approx {\bf A}_{ \lambda_y} = 
\sqrt{\frac{ \hbar }{2 \omega_{\boldsymbol{k}}\epsilon_0(2 \pi)^3}}
 \boldsymbol{e}_{y}$ as a result the absorption operator $R_{\lambda_y}$ is also independent on momentum $\boldsymbol{k}$. Then, the two-photon ionization amplitude for entangled photons, 
\(\tau_{{\rm ent};i\mu_f}^{(2)}\left( \boldsymbol{p}_f \right)\), 
can be expressed in terms of the corresponding amplitude for separable photons, 
\(\tau_{{\rm sep};i\mu_f}^{(2)}\left( \boldsymbol{p}_f \right)\),
\begin{align}
\tau_{{\rm ent};i\mu_f}^{(2)}\left( \boldsymbol{p}_f \right) &\approx -\frac{  8\pi e^2 c^2 i}{\hbar^2}   \frac{1}{i\Gamma_n}\sum_n  \left\langle\varphi_{f}\left|\boldsymbol{\alpha} \cdot {\bf A}_{ \lambda_y}\right| \varphi_n\right\rangle\left\langle\varphi_n\left|\boldsymbol{\alpha} \cdot {\bf A}_{ \lambda_y}\right| \varphi_{i}\right\rangle \int_{-\infty}^{\infty}\int_{-\infty}^{\infty} d \boldsymbol{k}_i d \boldsymbol{k}_s F_{ent}\left(\boldsymbol{k}_i,\boldsymbol{k}_s\right) = \frac{\tau_{{\rm sep};i\mu_f}^{(2)}\left( \boldsymbol{p}_f \right)\int_{-\infty}^{\infty}\int_{-\infty}^{\infty} d \boldsymbol{k}_i d \boldsymbol{k}_sF_{ent}\left(\boldsymbol{k}_i,\boldsymbol{k}_s\right)}{\int_{-\infty}^{\infty}\int_{-\infty}^{\infty} d \boldsymbol{k}_i d \boldsymbol{k}_s F_{sep}\left(\boldsymbol{k}_i,\boldsymbol{k}_s\right)}.
\label{tau_dipole}
\end{align}
Using \eqref{tau_dipole} together with \eqref{flux_sm} and \eqref{sigma_2}, we obtain the enhancement ratio as
\begin{align}
    \mathcal{R}^{(2)} &=
 \frac{V_{ent}}{V_{sep}} \frac{f_{ent}^{(2)}}{f_{sep}^{(2)}},
\end{align}
where $f_{ent/sep}^{(2)} = \frac{\left|\int_{-\infty}^{\infty}\int_{-\infty}^{\infty} d \boldsymbol{k}_i d \boldsymbol{k}_s F_{ent/sep}(\boldsymbol{k}_i,\boldsymbol{k}_s)\right|^2}{ \int_{-\infty}^{\infty}\int_{-\infty}^{\infty} d \boldsymbol{k}_i d \boldsymbol{k}_s 
|F_{ent/sep}(\boldsymbol{k}_i,\boldsymbol{k}_s)|^2\left( \frac{k_{i,z}}{k_i}+\frac{k_{s,z}}{k_s} \right)}$.
\\
\indent
Similarly, one can extend the enhancement ratio to the case of absorption of $n$-entangled photons ($n$ is even), 
\begin{align}
\tau^{(n)}_{ent}  = \frac{\tau^{(n)}_{sep}\int_{-\infty}^{\infty} d \boldsymbol{k}_1 \cdots d \boldsymbol{k}_nF_{ent}\left(\boldsymbol{k}_1,\cdots,\boldsymbol{k}_n\right)}{\int_{-\infty}^{\infty} d \boldsymbol{k}_1 \cdots d \boldsymbol{k}_n F_{sep}\left(\boldsymbol{k}_1,\cdots,\boldsymbol{k}_n\right)},
\end{align}
in which we assumed that $\psi_i = \int_{-\infty}^{\infty} d \boldsymbol{k}_1 \cdots d \boldsymbol{k}_nF_{ent}\left(\boldsymbol{k}_1,\cdots,\boldsymbol{k}_n\right)a^\dagger_{\lambda}(\boldsymbol{k}_1)\cdots a^\dagger_{\lambda}(\boldsymbol{k}_n)$ and that $F_{ent}\left(\boldsymbol{k}_1,\cdots,\boldsymbol{k}_n\right)$ is symmetric to any permutations of $\boldsymbol{k}_n$ and that polarization of the photons is independent on the direction of the vectors $\boldsymbol{k}_n$ (which is true for the case of paraxial approximation or the case when the wavevectors are confined in $2d$ plane). Then the flux is given by
\begin{align}
    \phi_{ent/sep} &=  \frac{2c}{V_{ent/sep}}\int_{-\infty}^{\infty} d \boldsymbol{k}_1 \cdots d \boldsymbol{k}_n \left| F\left(\boldsymbol{k}_1,\cdots,\boldsymbol{k}_n\right) \right|^2 \left( \frac{k_{1,z}}{k_1}+\cdots+\frac{k_{n,z}}{k_n} \right), 
\end{align}
and the enhancement ratio $\mathcal{R}^{(n)}$ is given by
\begin{align}
    \mathcal{R}^{(n)} &=
 \frac{V_{ent}}{V_{sep}} \frac{f_{ent}^{(n)}}{f_{sep}^{(n)}},
\end{align}
where $f_{ent/sep}^{(n)} = \frac{\left|\int_{-\infty}^{\infty} d \boldsymbol{k}_1 \cdots d \boldsymbol{k}_n F_{ent/sep}(\boldsymbol{k}_i,\cdots,\boldsymbol{k}_n)\right|^2}{ \int_{-\infty}^{\infty} d \boldsymbol{k}_1,\cdots, d \boldsymbol{k}_n 
|F_{ent/sep}(\boldsymbol{k}_1, \cdots,\boldsymbol{k}_n)|^2\left( \frac{k_{1,z}}{k_1}+\cdots+\frac{k_{n,z}}{k_n} \right)}$. 
\\
\indent
In Fig. \ref{Fig_SM} we plot the enhancement ratio $\mathcal{R}^{(2)}$ for the case of the ionization of a sodium atom in SAE approximation through the intermediate $4p_{3/2}$ level. The curves are obtained from Eq.~\eqref{R_dip}, while the discrete markers represent the numerical evaluation of Eq.~\eqref{R}, showing excellent agreement between analytical and numerical results.
\section{C. The two-photon amplitude}
\label{sec:two_ph_ampl}
In the current investigation, we assume that entangled photons are generated via the SPDC process and the twin state $\ket{\psi_{ent}}$ is given by \cite{Vega2022},
\begin{align}
\ket{\psi_{ent}} \propto \int_{-T_{int}}^{T_{int}} d t \int_{V_{crystal}} d \boldsymbol{r} \sum_{\alpha, \beta, \gamma = x,y,z}\chi_{\alpha \beta \gamma}^{(2)}(\boldsymbol{r}) E_{\mathrm{P}, \gamma}(\boldsymbol{r}, t) \hat{E}_\alpha^{(-)}(\boldsymbol{r}, t)\hat{E}_\beta^{(-)}(\boldsymbol{r}, t)|0,0\rangle,
\end{align}
where $\chi_{\alpha \beta \gamma}^{(2)}$ is the second-order nonlinear susceptibility tensor of the
crystal. We integrate over the interaction volume of the crystal $V_{crystal}$ and interaction time $T_{int}$, with $E_{\mathrm{p}, \gamma}(\boldsymbol{r}, t)$ being $\gamma$th component of the pump field which we assume to be linearly polarized in $y$ direction and propagate along positive $z$ axis
\begin{align}
 E_{p}(\boldsymbol{r}, t) = \int_{-\infty}^{\infty} d\boldsymbol{k}_p e^{-\frac{\Omega_p^2 k^2_{p,x}}{2}
-\frac{\Omega_{p,y}^2 k^2_{p,y}}{2}}e^{i\boldsymbol{k}_p \cdot \boldsymbol{r} - i\omega_pt}
 \theta \left( k_{p,z}\right)
 \boldsymbol{e}_y,
\end{align}
where $\Omega_p$ and $\Omega_{p,y}$ characterizing a spread of the components $k_{p,x}$ and $k_{p,y}$ of the momentum $\boldsymbol{k}_p$, respectively. The $\theta \left( k_{P,z}\right)$ is Heaviside function, which makes photons propagate in the positive $z$-direction, and $\boldsymbol{e}_y$ is the polarization vector in the $y$-direction. The quantized electric field $\hat{\boldsymbol{E}}^{(-)}$ which corresponds to idler and signal is given by
\begin{align}
\hat{\boldsymbol{E}}^{(-)}(\boldsymbol{r}, t)= & \frac{-i}{(2 \pi)^{3 / 2}} \sum_{\lambda=1,2} \int_{-\infty}^{\infty} d \boldsymbol{k}\left(\frac{\hbar \omega_{\boldsymbol{k}}}{2 \varepsilon_0}\right)^{1 / 2} \hat{a}^{\dagger}_{\lambda}(\boldsymbol{k})  e^{-i(\boldsymbol{k} \cdot \boldsymbol{r}-\omega t)}\theta \left( k_{z}\right) \boldsymbol{e}(\boldsymbol{k}, \lambda)
\end{align}
where $\hbar$ is the reduced Planck constant, $\epsilon_0$ is the vacuum permittivity, $\hat{a}_{\lambda}(\boldsymbol{k})$ and $\hat{a}^{\dagger}_{\lambda}(\boldsymbol{k})$ are the creation and annihilation operators of a photon with momentum $\boldsymbol{k}$ and helicity $\lambda$, respectively, such as
\begin{equation}
\left[\hat{a}_{\lambda}(\boldsymbol{k}), \hat{a}^{\dagger}_{\lambda'}(\boldsymbol{k}')\right]=\delta^{(3)}\left(\boldsymbol{k}-\boldsymbol{k}'\right) \delta_{\lambda \lambda'}.
\end{equation}
The energy of the photon is $\omega_{\boldsymbol{k}}=c|\boldsymbol{k}|$ and $\boldsymbol{e}(\boldsymbol{k},\lambda)$ is the transversal dimensionless polarization vector that obeys $\boldsymbol{k} \cdot \boldsymbol{e}(\boldsymbol{k},\lambda)=$ $\boldsymbol{e}(\boldsymbol{k},1)^* \cdot \boldsymbol{e}(\boldsymbol{k},2)=0$.

The two-photon state can thus be written as
\begin{align}
|\psi_{ent}\rangle & \propto \int_{-\infty}^{+\infty} d t \int_{-\infty}^{+\infty} d x 
\int_{-\infty}^{+\infty} d y
\int_{-L / 2}^{+L / 2} d z \sum_{\alpha, \beta=x,y,z} \chi_{\alpha \beta y}^{(2)}(x,y,z)  \int_{-\infty}^{+\infty} d\boldsymbol{k}_p e^{-\frac{\Omega_p^2 k^2_{p,x}}{2}
-\frac{\Omega_{p,y}^2 k^2_{p,y}}{2}}e^{i\left(k_{p,x}x+k_{p,y}y+k_{p,z}z - \omega_p t \right) }
 \theta \left( k_{P,z}\right)
  \notag \\ 
&\sum_{\lambda_i} \int_{-\infty}^{\infty} d \boldsymbol{k}_i \omega_{\boldsymbol{k}_i}^{1/2} e^{-i(k_{i,x}x+k_{i,y}y+k_{i,z}z-\omega_i t)}\theta \left( k_{i,z}\right) e_{\alpha}(\boldsymbol{k}_i, \lambda_i) \hat{a}^{\dagger}_{\lambda_i}(\boldsymbol{k}_i) \sum_{\lambda_s} \int_{-\infty}^{\infty} d \boldsymbol{k}_s \omega_{\boldsymbol{k}_s}^{1/2} e^{-i(k_{s,x}x+k_{s,y}y+k_{s,z}z-\omega_s t)}\theta \left( k_{s,z}\right) e_{\beta}(\boldsymbol{k}_s, \lambda_s) \hat{a}^{\dagger}_{\lambda_s}(\boldsymbol{k}_s)|0,0\rangle. \label{eq:GerneralTwo}
\end{align}
To take the spatial and time integrals, we use the following relations
\begin{align}
&\int_{-\infty}^{+\infty} d x e^{i\left(k_{p,x}-k_{s,x} - k_{i,x}\right) x} \propto \delta\left(k_{p,x}-k_{s,x} - k_{i,x}\right), \\
&\int_{-\infty}^{+\infty} d y e^{i\left(k_{p,y}-k_{s,y} - k_{i,y}\right) y} \propto \delta\left(k_{p,y}-k_{s,y} - k_{i,y}\right), \\
&\int_{-L / 2}^{+L / 2} d z e^{i\left(k_{p,z}-k_{s,z} - k_{i,z}\right) z} \propto \operatorname{sinc}\left(\frac{\Delta k_z L}{2}\right), \\
&\int_{-\infty}^{+\infty} d t e^{i\left(\omega_s+\omega_i-\omega_p\right) t} \propto \delta\left(\omega_s+\omega_i-\omega_p\right) \propto \delta\left(\omega_s/c+\omega_i/c-k_p\right),
\end{align}
in which $\Delta k_z = k_{P,z}-k_{s,z} - k_{i,z}$ and consequently, Eq. \eqref{eq:GerneralTwo} reduces to
\begin{align}
|\psi_{ent}\rangle & \propto  
\sum_{\alpha, \beta=x,y,z} \chi_{\alpha \beta y}^{(2)}  \int_{-\infty}^{+\infty} \int_{-\infty}^{\infty} \int_{-\infty}^{\infty} d \boldsymbol{k}_s d \boldsymbol{k}_i d\boldsymbol{k}_p \delta\left(k_{p,x}-k_{s,x} - k_{i,x}\right)  \delta\left(k_{p,y}-k_{s,y} - k_{i,y}\right)  \operatorname{sinc}\left(\frac{\Delta k_z L}{2}\right)  \delta\left(k_s+k_i-k_p\right) e^{-\frac{\Omega_p^2 k^2_{p,x}}{2}
-\frac{\Omega_{p,y}^2 k^2_{p,y}}{2}}
 \theta \left( k_{p,z}\right)
  \notag \\
&\sum_{\lambda_i} \omega_{\boldsymbol{k}_i}^{1/2} \theta \left( k_{i,z}\right) e_{\alpha}(\boldsymbol{k}_i, \lambda_i) \hat{a}^{\dagger}_{\lambda_i}(\boldsymbol{k}_i) \sum_{\lambda_s}  \omega_{\boldsymbol{k}_s}^{1/2} \theta \left( k_{s,z}\right) e_{\beta}(\boldsymbol{k}_s, \lambda_s) \hat{a}^{\dagger}_{\lambda_s}(\boldsymbol{k}_s)|0,0\rangle,
\end{align}
where we have assumed that the crystal is uniform without poling (making $\chi^{(2)}$ independent of $x,y,z$). In the next step, we use the following identity for the delta function, 
\begin{align}
\delta\left(\omega_s/c+\omega_i/c-k_p\right) = \frac{\delta\left(k_{p,z} - \sqrt{(k_s+k_i)^2-k_{p,x}^2-k_{p,y}^2}\right)(\omega_s+\omega_i)}{\sqrt{(k_s+k_i)^2-k_{p,x}^2-k_{p,y}^2}c}, 
\end{align}
where we took into account that $k_{p,z}>0$.  Using the above delta function, the twin state can be simplified further to 
\begin{align}
|\psi_{ent}\rangle & \propto    
\sum_{\alpha, \beta=x,y,z} \chi_{\alpha \beta y}^{(2)} \int_{-\infty}^{\infty} \int_{-\infty}^{\infty} d \boldsymbol{k}_s d\boldsymbol{k}_i \frac{(\omega_s+\omega_i)}{\sqrt{(k_s+k_i)^2-k_{p,x}^2-k_{p,y}^2}} 
  \operatorname{sinc}\left(\frac{\Delta k_z L}{2}\right) e^{-\frac{\Omega_p^2 k^2_{p,x}}{2}
-\frac{\Omega_{p,y}^2 k^2_{p,y}}{2}}
\notag  \\
&\sum_{\lambda_i}  \omega_{\boldsymbol{k}_i}^{1/2} \theta \left( k_{i,z}\right) e_{\alpha}(\boldsymbol{k}_i, \lambda_i) \hat{a}^{\dagger}_{\lambda_i}(\boldsymbol{k}_i) \sum_{\lambda_s}  \omega_{\boldsymbol{k}_s}^{1/2} \theta \left( k_{s,z}\right) e_{\beta}(\boldsymbol{k}_s, \lambda_s) \hat{a}^{\dagger}_{\lambda_s}(\boldsymbol{k}_s)|0,0\rangle,
\end{align}
where $k_{p,x} = k_{s,x}+k_{i,x}$, $k_{p,y} = k_{s,y}+k_{i,y}$ and $k_{p,z} = \sqrt{(k_s+k_i)^2-k_{p,x}^2-k_{p,y}^2}$.
Considering that we want to have a fixed frequency and $k_y \approx 0$, we apply the filtering to the signal and idler. This means that we should add a corresponding filter function $g = e^{-\frac{k_{s/i,y}^2\Omega_y^2}{2}} e^{-\frac{(\omega_{s/i} - \omega)^2T^2}{2}}$ where first exponent filters out photons with $k_y \approx 0$ and the second one filters out single frequency $\omega$ for sufficiently long pulse duration $T$. Since $k_y \approx 0$ for both signal and idler, then in summation over polarization $\lambda_{i/s}$, one can take one of the polarizations being along $y$ and another perpendicular. We also assume that $\chi^{(2)}_{yyy} = const$ is the largest contribution, which is true for most crystals \cite{Vega2022}. Using the vacuum dispersion relationship for signal and idler photons, we get 
\begin{align}
|\psi_{ent}\rangle & \propto \int_{-\infty}^{\infty}\int_{-\infty}^{\infty} d \boldsymbol{k}_i d \boldsymbol{k}_s 
  \frac{\sqrt{\omega_{\boldsymbol{k}_s}\omega_{\boldsymbol{k}_i}}(k_s+k_i)}{\sqrt{(k_s+k_i)^2-k_{p,x}^2-k_{p,y}^2}}\operatorname{sinc}\left(\frac{\Delta k_z L}{2}\right)  
  e^{-\frac{\Omega_p^2 k^2_{p,x}}{2}
-\frac{\Omega_{p,y}^2 k^2_{p,y}}{2}}
   e^{-\frac{\left(k_{s,y}^2+k_{i,y}^2\right)\Omega_y^2}{2}} e^{-\frac{\left((\omega_{s} - \omega)^2+(\omega_{i} - \omega)^2\right)T^2}{2}}
   \theta \left( k_{i,z}\right) 
 \theta \left( k_{s,z}\right) 
    \notag \\
&   
\hat{a}^{\dagger}_{\lambda_i}(\boldsymbol{k}_i) 
\hat{a}^{\dagger}_{\lambda_y}(\boldsymbol{k}_s)|0,0\rangle.
\end{align}
The use of the vacuum dispersion relation eliminates the need to account for crystal-specific refractive index variations, which would require detailed knowledge of the material dispersion and reduce the generality of the analysis \cite{Vega2022}. Since only photons with transverse wave vectors smaller than the free-space wave vector can exit the crystal, the accessible mode range is fundamentally determined by the vacuum dispersion condition. Deviations from this condition due to the actual crystal dispersion affect only the detailed shape of the phase-matching function, without altering the accessible transverse momentum range (especially in the case of a very thin crystal) or the main conclusions of this work.

Next, since we have small $k_{p}$ and $k_{p,y}$ compared to $k_p = k_i+k_s$, then $\frac{(k_s+k_i)}{\sqrt{(k_s+k_i)^2-k_{p,x}^2-k_{p,y}^2}}\approx 1$ and due to the beam being narrowband, we have $\omega_i\approx\omega_s\approx \omega$, then twin state  can be written as 
\begin{align}
|\psi_{ent}\rangle & =  
  \int_{-\infty}^{\infty}\int_{-\infty}^{\infty} d \boldsymbol{k}_i d \boldsymbol{k}_s F_{ent}(\boldsymbol{k}_i,\boldsymbol{k}_s) \hat{a}^{\dagger}_{\lambda_i}(\boldsymbol{k}_i)     \hat{a}^{\dagger}_{\lambda_y}(\boldsymbol{k}_s)|0,0\rangle,
\end{align}
where $F_{ent}(\boldsymbol{k}_i,\boldsymbol{k}_s) = \frac{C_{ent}}{\sqrt{2}}\tilde{F}_{ent}(\boldsymbol{k}_i,\boldsymbol{k}_s)$ with $C_{ent}$ being a normalization factor and $\tilde{F}_{ent}(\boldsymbol{k}_i,\boldsymbol{k}_s)$
\begin{align}
\tilde{F}_{ent}(\boldsymbol{k}_i,\boldsymbol{k}_s) & =  
  \operatorname{sinc}\left(\frac{\Delta k_z L}{2}\right) 
  E_p(\boldsymbol{k}_p,\Omega_p)
   e^{-\frac{\left(k_{s,y}^2+k_{i,y}^2\right)\Omega_y^2}{2}} e^{-\frac{\left((\omega_{s} - \omega)^2+(\omega_{i} - \omega)^2\right)T^2}{2}}
   \theta \left( k_{i,z}\right) 
 \theta \left( k_{s,z}\right). \label{eq.FENT}
\end{align}
Here $\Delta k_z = k_{p,z} - k_{s,z} - k_{i,z} = \sqrt{(k_s+k_i)^2 - k^2_{p,x}-k^2_{p,y}} - k_{s,z} - k_{i,z} = \sqrt{(k_s+k_i)^2 - 
(k_{s,x}+k_{i,x})^2-(k_{s,y}+k_{i,y})^2} - k_{s,z} - k_{i,z} \approx \sqrt{(k_s+k_i)^2 - 
(k_{s,x}+k_{i,x})^2} - k_{s,z} - k_{i,z}$ and $k_{s/i,z} \approx \sqrt{k^2_{s/i}-k^2_{s/i,x}}$, and $E_p(\boldsymbol{k}_p,\Omega_p) = e^{-\frac{\Omega_p^2 k^2_{p,x}}{2}
-\frac{\Omega_{p,y}^2 k^2_{p,y}}{2}}$.
For the case of separable photons, we use the following expression for the two-photon amplitude
\begin{align}
|\psi_{sep}\rangle & =\int_{-\infty}^{\infty}\int_{-\infty}^{\infty} d \boldsymbol{k}_i d \boldsymbol{k}_s  F_{sep}(\boldsymbol{k}_i, \boldsymbol{k}_f) 
\hat{a}^{\dagger}_{\lambda_i}(\boldsymbol{k}_i)    \hat{a}^{\dagger}_{\lambda_y}(\boldsymbol{k}_s)|0,0\rangle,   \label{eq.FSEP}
\end{align}
where $F_{sep}(\boldsymbol{k}_i, \boldsymbol{k}_f) = \frac{C_{sep}}{\sqrt{2}}\tilde{F}_{sep}(\boldsymbol{k}_i, \boldsymbol{k}_f)$ with $C_{sep}$ being a normalization constant and $\tilde{F}_{sep}(\boldsymbol{k}_i, \boldsymbol{k}_f)$ is given by
\begin{align}
\tilde{F}_{sep}(\boldsymbol{k}_i, \boldsymbol{k}_f) & =  e^{-\frac{\left(k_{s,x}^2+k_{i,x}^2\right)\Omega_p^2}{2}} 
e^{-\frac{\left(k_{s,y}^2+k_{i,y}^2\right)\Omega_y^2}{2}} e^{-\frac{\left((\omega_{s} - \omega)^2+(\omega_{i} - \omega)^2\right)T^2}{2}}   \theta \left( k_{i,z}\right) 
\theta \left( k_{s,z}\right). 
\label{eq.FSEP}
\end{align}
\end{widetext}

\end{document}